\documentclass[graybox]{svmult}

\usepackage{mathptmx}       
\usepackage{helvet}         
\usepackage{courier}        
\usepackage{type1cm}        
\usepackage{makeidx}         
\usepackage{graphicx}        
\usepackage{multicol}        
\usepackage[bottom]{footmisc}

\newcommand{\hMpc}{{\ifmmode{h^{-1}{\rm Mpc}}\else{$h^{-1}$Mpc}\fi}}
\newcommand{\hkpc}{{\ifmmode{h^{-1}{\rm kpc}}\else{$h^{-1}$kpc}\fi}}
\newcommand{\hMsun}{{\ifmmode{h^{-1}{\rm {M_{\odot}}}}\else{$h^{-1}{\rm{M_{\odot}}}$}\fi}}
\newcommand{\kms}{{\,\rm km\,s^{-1}}} 
\newcommand{\keV}{\>{\rm keV}}
\def\lesssim{_ <\atop{^\sim}}

\makeindex             

\begin{document}

\title*{Constrained Local UniversE Simulations (CLUES) }
\author{Stefan Gottl\"ober,   Yehuda Hoffman and Gustavo Yepes}
\institute{Stefan Gottl\"ober \at 
Astrophysical Institute Potsdam, An der Sternwarte 16, 14482 Potsdam, Germany
\email{sgottloeber@aip.de}
\and Yehuda Hoffman \at 
Racah Inst. of Physics
Hebrew University
Jerusalem 91904, Israel
\email{hoffman@huji.ac.il}
\and Gustavo Yepes
Departamento de F\'\i sica Teorica C-15; 
Universidad Autonoma de Madrid;
Madrid 28049, Spain
\email{gustavo.yepes@uam.es}}
%
%
\maketitle

\abstract*{ The local universe is the best known part of our
universe. Within the CLUES project ({\tt http://clues-project.org} ---
Constrained Local UniversE Simulations) we perform numerical
simulations of the evolution of the local universe. For these
simulations we construct initial conditions based on observational
data of the galaxy distribution in the local universe. Here we review
the technique of these constrained simulations. In the second part we
summarize our predictions of a possible Warm Dark Matter cosmology for
the observed local distribution of galaxies and the local spectrum of
mini-voids as well as a study of the satellite dynamics in a simulated
Local Group.  }

\abstract{ The local universe is the best known part of our
universe. Within the CLUES project ({\tt http://clues-project.org} ---
Constrained Local UniversE Simulations) we perform numerical
simulations of the evolution of the local universe. For these
simulations we construct initial conditions based on observational
data of the galaxy distribution in the local universe. Here we review
the technique of these constrained simulations. In the second part we
summarize our predictions of a possible Warm Dark Matter cosmology for
the observed local distribution of galaxies and the local spectrum of
mini-voids as well as a study of the satellite dynamics in a simulated
Local Group.  }

\section{Introduction}
\label{sec:intro_h009z}

During the last decade cosmological parameters have been determined to
a precision of just a few percent giving rise to the standard model of
cosmology: a flat Friedmann universe whose mass-energy content is
dominated by a cosmological constant (the $\Lambda$ term), a cold dark
matter (CDM) component and baryons. The convergence to a standard
model of cosmology sets the framework for studying the formation of
both the large scale structure and of galaxies within that model. The
main goal of this study is to achieve a physical understanding of how
the observed structure emerged within the context of the given
cosmological model.  This model specifies the cosmological expansion
history, the initial conditions and the material content of the
universe. These plus the knowledge of the physical laws governing the
dynamics of the dark matter, baryons and radiation provide the
framework within which a successful model of galaxy formation can be
developed.

The basic paradigm of structure formation in the $\Lambda$CDM was
formulated more than 30 years ago\cite{whiterees78}.  It suggests
that dark matter (DM) clusters to form DM halos, within which galaxy
formation takes place via complex baryonic physics.  Intensive
numerical efforts of the last decade have validated the basic White \&
Rees picture.  Yet, the process of galaxy formation is much more
complicated than that simple picture.  DM halos grows via process of
mergers of substructures and galaxy formation proceeds by the combined
action of clumpy and anisotropic gas accretion and mergers dwarf
galaxies. As dwarf galaxies cross the virial radius of the DM halo
they become satellites, subject to tidal forces exerted by the central
potential.  This transforms the satellites into tidal streams and
eventually phase mixing dissolves the streams into the stellar halo of
galaxies. This process manifests itself in the combined color and
dynamical phase space of galactic stella halos.  This is the universal
mode of galaxy formation but it can be observed and analyzed only in
the very local universe, resulting in the so-called near-field
cosmology. This motivates cosmologists to turn their attention and
study the archeology of the Local Group (LG) in their quest for
understanding galaxy formation. This also motivates us to simulate the
formation of the LG in the most realistic possible way.

Cosmological simulations may cover a large dynamical and mass range.
A representative volume of the universe should be large, but this
comes at the expense of the mass resolution, which must be decreased
in proportion to the simulated volume size.  One may reduce the box
size at the expense of possibly not being representative. To overcome
this problem we use smaller simulation boxes specifically designed to
represent the observed local universe.  The algorithm of constrained
realizations of Gaussian field provides a very attractive method of
imposing observational data as constraints on the initial conditions
and thereby yielding structures which can closely mimic those in the
actual universe.  The prime motivation of the CLUES project is to
construct simulations that reproduce the local cosmic web and its key
'players', such as the Local Supercluster, the Virgo cluster, the Coma
cluster, the Great Attractor and the Perseus-Pisces supercluster.  The
main drawback of current constrained simulations is that they do not
directly constrain the sub-megaparsec scale structure, yet they enable
the simulation of these scales within the correct environment.  Such
simulations provide a very attractive possibility of simulating
objects with properties close to the observed LG and situated within
the correct environment.  Their random origin, however, limits the
predictive power of the simulations and in particular the constraining
power of our 'near field cosmology' experiments. It is one of our main
goals to improve the constraining power and ability of the simulations
to reproduce the actual observed LG.

\section{Constrained Simulations}
\label{sec:CR_h009z}

During the past  few years we have performed a series of constrained
simulations of the local universe. We continue this research to answer
within the next generation of constrained simulations the question how
unique the LG  is. To this end we will use more and better
observational data of the local Universe. Our main motivation is to
use the constrained simulations as a numerical near-field cosmological
laboratory for experimenting with the complex DM and gastrophysics
processes.

\subsection{ Observational Data}
\label{subsec:obs_h009z}

Observational data of the nearby universe is used as constraints on
the initial conditions and thereby the resulting simulations reproduce
the observed large scale structure.  The implementation of the
algorithm of constraining Gaussian random fields
\cite{hoffmanribak:91} to observational data and a description of the
construction of constrained simulations was described already
elsewhere \cite{kravtsovetal:02,klypinetal:03}. Here we briefly
describe the observational data used.  Two different observational
data is used to set up the initial conditions. The first is made of
radial velocities of galaxies drawn from the MARK III
\cite{willicketal:97}, SBF \cite{tonryetal:01} and the
Karachentsev\cite{karachentsev:04} catalogs. Peculiar velocities are
less affected by non-linear effects and are used as constraints as if
they were linear quantities \cite{zaroubietal:99}.  The other
constraints are obtained from the catalog of nearby X-ray selected
clusters of galaxies \cite{reiprichetal:02}. Given the virial
parameters of a cluster and assuming the spherical top-hat model one
can derive the linear overdensity of the cluster. The estimated linear
overdensity is imposed on the mass scale of the cluster as a
constraint.  For the CDM cosmogonies the data used here constrains the
simulations on scales larger than $\approx 5 \hMpc$
\cite{klypinetal:03}. It follows that the main features that
characterise the local universe, such as the Local Supercluster, Virgo
cluster, Coma cluster and Great attractor, are all reproduced by the
simulations. The small scale structure is hardly affected by the
constraints and is essentially random.

\subsection{Constrained Initial conditions}
\label{subsec:IC_h009z}

The Hoffman-Ribak algorithm \cite{hoffmanribak:91} is used to generate
the initial conditions as constrained realizations of Gaussian random
fields on a $256^3$ uniform mesh, from the observational data
mentioned above.  Since these data only constrain scales larger than a
very few Mpc, we need to perform a series of different realizations in
order to obtain one which contain an LG candidate with the correct
properties (e.g. two halos with proper position relative to
each-other, mass, negative radial velocity, etc).  Hence the low
resolution, $256^3$ particle simulations were searched and the one
with the most suitable Local Group like objects were chosen for follow up, high
resolution re-simulations.  We generated more than 200 realizations of
these simulations and all of them were evolved since the starting
redshift ($z=50$) till present time ($z=0$) using the TREEPM N-body
algorithm in GADGET2.  We used 64 processors of HLRB2 Altix
supercomputer and the typical wall clock time for each of them was of
the order of 10 hours.

High resolution extension of the low resolution constrained
realizations were then obtained by creating an unconstrained
realization at the desired resolution, FFT-transforming it to k-space
and substituting the unconstrained low k modes with the constrained
ones. The resulting realization is made of unconstrained high k modes
and constrained low k ones.  In order to be able to zoom into the
desired Local Group object, we set up the unconstrained realization to
the maximum number of particles we can allocate in one node of the
HLRB2 Altix cluster.  Since our initial conditions generator code is
OpenMP parallel, we can only run it in shared memory mode.  Thus,
thanks to the cc-NUMA architecture of the Altix, we can address as
much as 2 Tbytes of RAM for an OpenMP program.  This means we can
accommodate as many as $4096^3$ particles in this memory. We used up
to 500 processors in one node and 1.5 Tbytes of memory for that. The
typical cpu time per run is of the order of 5-6 hours wall clock. Most
of this time was taken to generate the sequence of Gaussian random
numbers, which was done in serial mode.  We would not have been able
to proceed in this project if we had not have access to this
computer. In fact, there are very few architectures in the world with
this huge amount of shared memory per node.  We are now rewriting the
initial condition generator code to MPI to overcome these limitations.

We  did two different kind of initial conditions using the procedure
described above. On one hand we were interested in simulating the
large scale structures of the whole simulation box of $64 \hMpc$ with
enough resolution to resolve  halos which can be related with the dwarf
galaxies  in our real Universe.  To this end, once we generated the
initial conditions for the largest number of particles, (i.e. $4096^3$)
then we degraded them  down to a maximum of $1024^3$ in the whole box. 

On the other hand, we were also interested in re-simulating with very
high resolution the formation of the LG candidates found in the low
resolution simulations. To this end, we re-simulate the evolution of
this region of interest, using the full resolution (equivalent to
$4096^3$ effective particles) only within a sphere of just $2
\hMpc$. Outside this region, the simulation box is populated with
lower resolution (i.e higher mass) particles. We were thus able to
achieve high resolution in the region of interest, while maintaining
the correct external environment. These initial conditions were set up
at very high redshift ($z=100$) to avoid spurious effects due to cell
crossing in the high resolution volume.  We then populate this volume
with two different species, dark matter and SPH gas particles, for the
case of running hydrodynamical simulations of these objects.

\subsection{Description of  Simulations}

Using the above initial conditions, we carried out the simulations
using the MPI N-body + SPH code GADGET2, developed by
V. Springel\cite{springel:05}.  The total amount of computing time
that has been invested in all the experiments done within the CLUES
collaboration has been tremendous.  We had to distribute the work
among different supercomputers in Europe, mainly the HLRB2 at LRZ, the
MareNostrum at BSC, and JUMP and more recently Juropa at J\"ulich.  In
all cases, the initial conditions were always generated in HLRB2 at
LRZ for the reasons explained above.

In total we have run  3 big collisionless (N-body only) 
simulations with 1  billion particles each (i.e. $1024^3$), and  2
more are currently running. Two of these simulations  correspond to
one realization with  WMAP3 cosmological parameters, that was done
both assuming CDM and WDM with $m_{WDM} = 1\keV $ and  the other one is
a  CDM realization with   WMAP5 cosmological parameters. 

The simulations started at $z=50$ and were evolved until $z=0$.  We
stored of the order of 190 snapshots, 60 of them with very fine time
interval of 15 million years between them, until $z=6$.  Then, we
enlarged the time interval to 100 million years between consecutive
snapshots. Since each snapshot is 32 Gbytes in size, we stored of the
order of 6.1 Tbytes of data in each run.  We used 500 MPI processors
and the typical computing time needed for these runs was of the order
of 130-150k CPU hours.

The other set of simulations we have also performed correspond to the
re-simulations of the LG object found in two of the constrained
realizations. As explained above, we have re-simulated a sphere with a
radius of $2 \hMpc$ around the object position at present time, with
the maximum resolution allowed by our initial condition setting
(i.e. $4096^3 $ effective particles in the re-simulate volume).  We
use the same set of initial conditions to run two simulations, one
with dark matter only and another one with dark matter, gas dynamics,
cooling, star formation and supernovae feedback.  In our SPH
simulation, each high resolution dark matter particle was replaced by
an equal mass, gas - dark matter particle pair with a mass ratio of
roughly 1:5.

Due to the strong clustering of the objects formed in these
simulations, the GADGET code does not scale very well with number of
processors. Thus, we decided to use the minimum necessary to allocate
all the data into memory. Total number of particles in these
simulations are of the order of 130-150 million.  On HLRB2 Altix at
LRZ we have used 64 to 128 processors for these runs.  The total
amount of cpu time exceeded the computational resources allocated in
normal projects, so we used also most of the time that was granted to
us from DEISA Extreme Computing Initiative (DECI) in two consecutive
projects named, SIMU-LU and SIMUGAL-LU. Each of them consistent of 1
million cpu hours divided among MareNostrum and HLRB2. We also
benefited from the DEISA high speed network in order to transfer the
snapshots between the two centres.

\paragraph{\bf Simulations of the full box}
\label{subsec:CRfullbox_h009z}

\begin{figure}[t]
\sidecaption[t]
\includegraphics[width=0.5\textwidth]{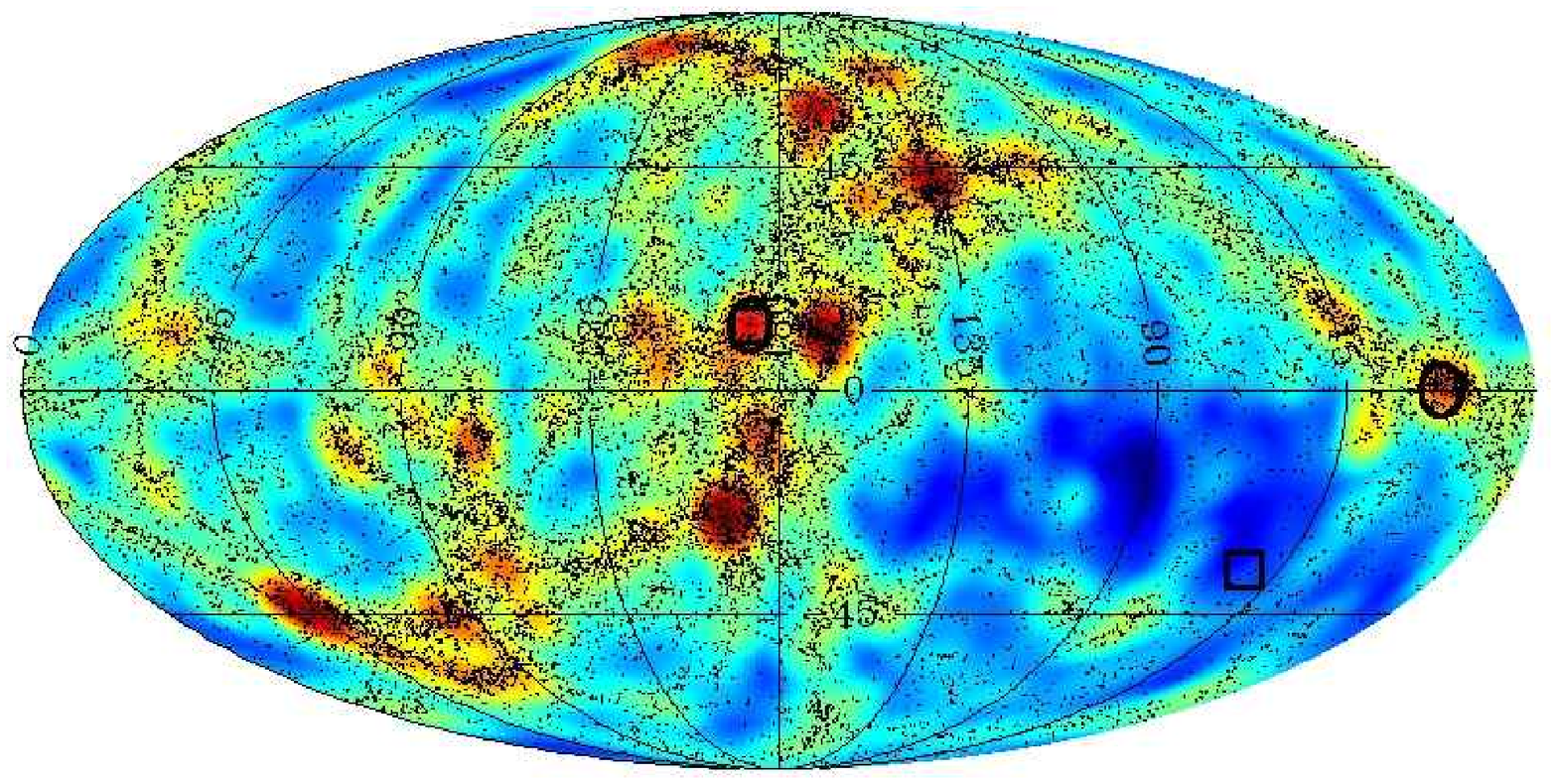}
\includegraphics[width=0.5\textwidth]{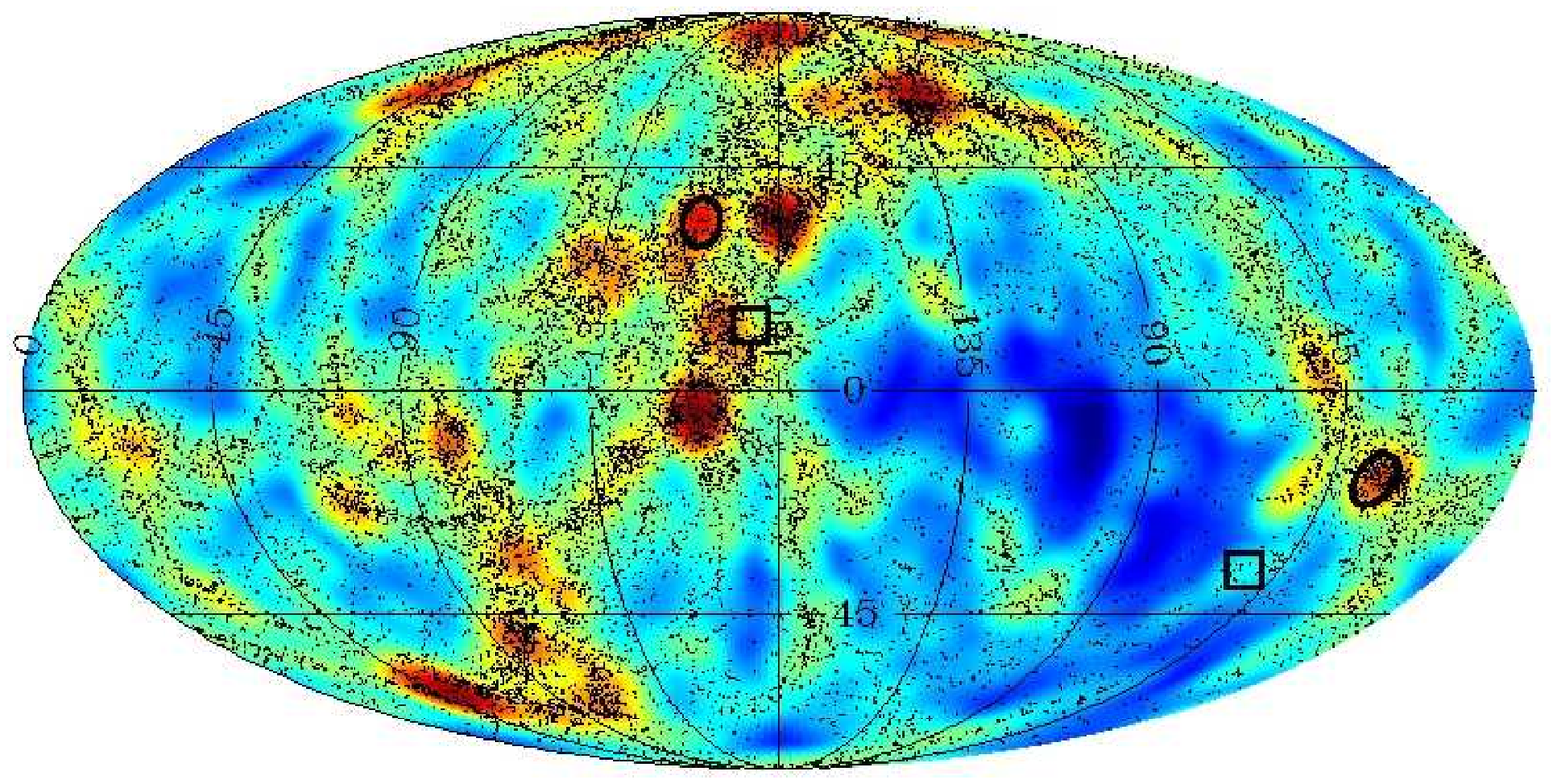}
\caption{The left  and the right plot show two projected sky maps with halos up to a distance of $20 \hMpc$ form the simulated Milky Way (see text for an explanation of the used coordinate system).}
\label{fig_h009z:proj_sky}
\end{figure}

To find a LG  candidate we first identify in the constrained
simulation the Virgo cluster. Then we search for an object which
closely resembles the Local Group and is in the right direction and
distance to Virgo.  Based on the locations of the LG and Virgo we now
define a coordinate system similar to the super-galactic coordinate
system.  We assume that the equatorial plane of the super-galactic 
coordinate system lies in the super-galactic plane, which is spanned by
the Local Group and the Local Supercluster. Thus we need another
object besides Virgo to define this plane. As shown by Zavala et al
\cite{zavalaetal:09} the Ursa Major cluster is a natural
choice. Rotating the coordinate system until the simulated Virgo
cluster is at the same angular position as the observed one we have
fixed our coordinate system. This coordinate system is visualised in
the left panel of Fig.~\ref{fig_h009z:proj_sky} which shows a sky map
with the angular distribution of halos within $20\hMpc$ from the
simulated Milky Way in equatorial coordinates (RA, DEC) in a Mollweide
projection. Here the value of each pixel is given by a mass-weighted
count of all halos located in that pixel (high density: red, low
density: blue) and all halos with masses larger than
$5\times10^{9}\hMsun$ appear as black points (see \cite{zavalaetal:09}
for details).

The simulated Local Supercluster can be seen as prominent vertical
structure in the left panel of Fig.~\ref{fig_h009z:proj_sky}. By
construction the simulated Virgo cluster (black circle) and the real
one (black square) are at the same position in the centre of the
plot. However, the simulated position (black circle on the right edge
of the plot) of a second nearby cluster, Fornax, differs from its
observed position (the black square). Such deviations are within the
expected variations of the constrained simulations. In fact, small
scale structure (as the position of the local group) are not
constrained by the observational data. The sky maps shown in
Fig.~\ref{fig_h009z:proj_sky} depend obviously of the chosen origin
(MW). The situation can be improved by an additional adjustment of the
coordinate system.  One can relax the requirement that the angular
position of the simulated and real Virgo has to coincide and use a
coordinate system that minimises the quadratic sum of the distances
between the simulated and real clusters Virgo and Fornax
(Fig.~\ref{fig_h009z:proj_sky}, right panel).

\paragraph{\bf Zoomed Simulations of the Local Group}
\label{subsec:CRLG_h009z}

\begin{figure}[t]
\sidecaption[t]                   
\includegraphics[width=\textwidth]{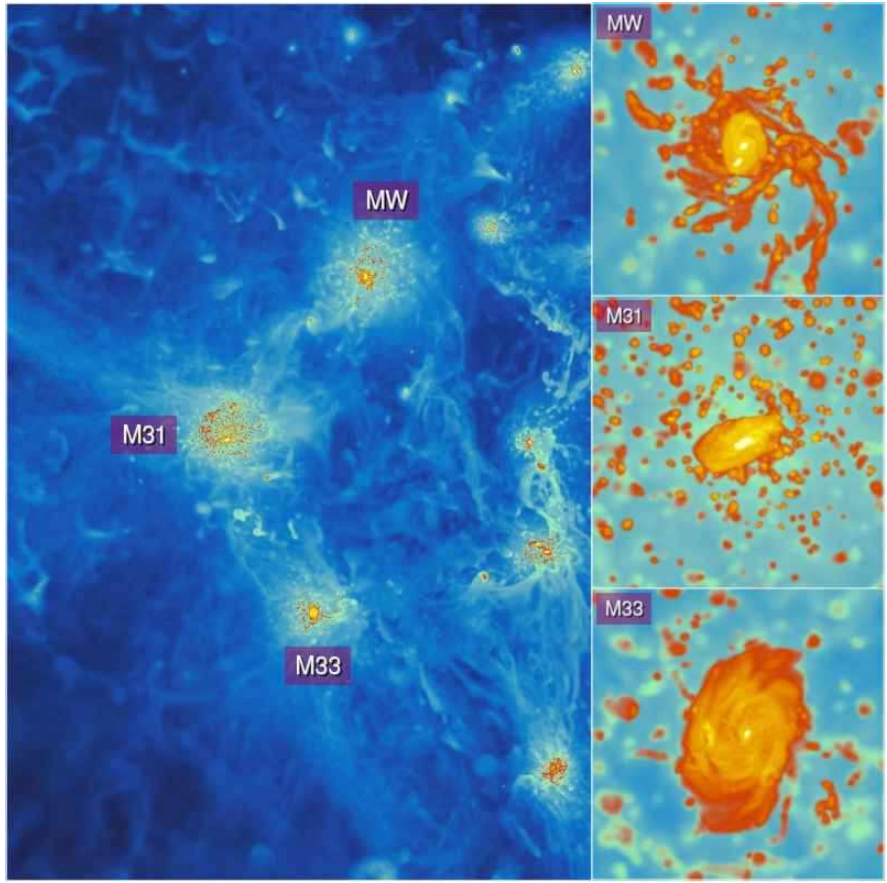}
\caption{The left hand side of the plot shows the gas distribution in the Local Group. On the right hand side we zoom  on the three objects (MW, M31, M33)}
\label{fig_h009z:LG_gas}
\end{figure}

On the left panel of Fig. \ref{fig_h009z:LG_gas} the gas distribution
in the Local Group is shown. The size of the plot is about $2 \hMpc$
across, viewed from a distance of $3.3 \hMpc$. On the three right
panels we show the gas disks of the three main galaxies as seen from a
distance of $250 \hkpc$, the size of the plot is about $50 \hkpc$. 
Since the three disks are seen from the
same distance and direction one can evaluate the relative orientation
of the three gas disks. Note, that the M31 and MW disks are smaller than
the disk of M33 due to major mergers which these objects had recently
($\sim z=0.6$). In the section  \ref{subsec:newCR_h009z} we have
described the new realizations of the local group which do not show
such recent major mergers.

The images were rendered with A. Khalatyan's SPMViewer using the
remote visualisation server RVS1 at LRZ. This allowed us to render the
particles directly on the GPU, which took one to two minutes per frame
when using all 52 million gas particles of the high resolution region.
We chose a logarithmic colour table for visualising the density
distribution of the gas: dark colours for low density and bright
colours for high density regions. The colours and density range are
adjusted such that the faint structures and filaments connecting the
three main galaxies become visible. For the zoomed images we shifted
the density range for colour mapping by a factor of $10^{0.5}$ in
order to improve the contrast and to to enhance the spiral arm
features of the gas disks.

\subsection{\bf An Ensemble of Constrained Simulations}
\label{subsec:newCR_h009z}

As we explained above, we generated an ensable of 200 constrained
simulations within the framework of the WMAP5 cosmological model.
These low resolution simulations have been inspected to find suitable
Local Group like objects.  Typically 10 such objects fulfil our
selection criteria\cite{martinezetal:09} in a volume of $(64~\hMpc)^3$.
However, these objects are not necessarily located close to the actual
location of the Local Group, namely nearby and in the right direction
of the the simulated Virgo cluster and the Local Supercluster. In
addition to the LG identified in the WMAP3 simulation (see
Fig. \ref{fig_h009z:LG_gas}) we have identified five more objects which
allow us to study the scatter in the properties of these ''Local
Groups''.

\begin{figure}[t]
\sidecaption[t]
\includegraphics[width=0.5\textwidth]{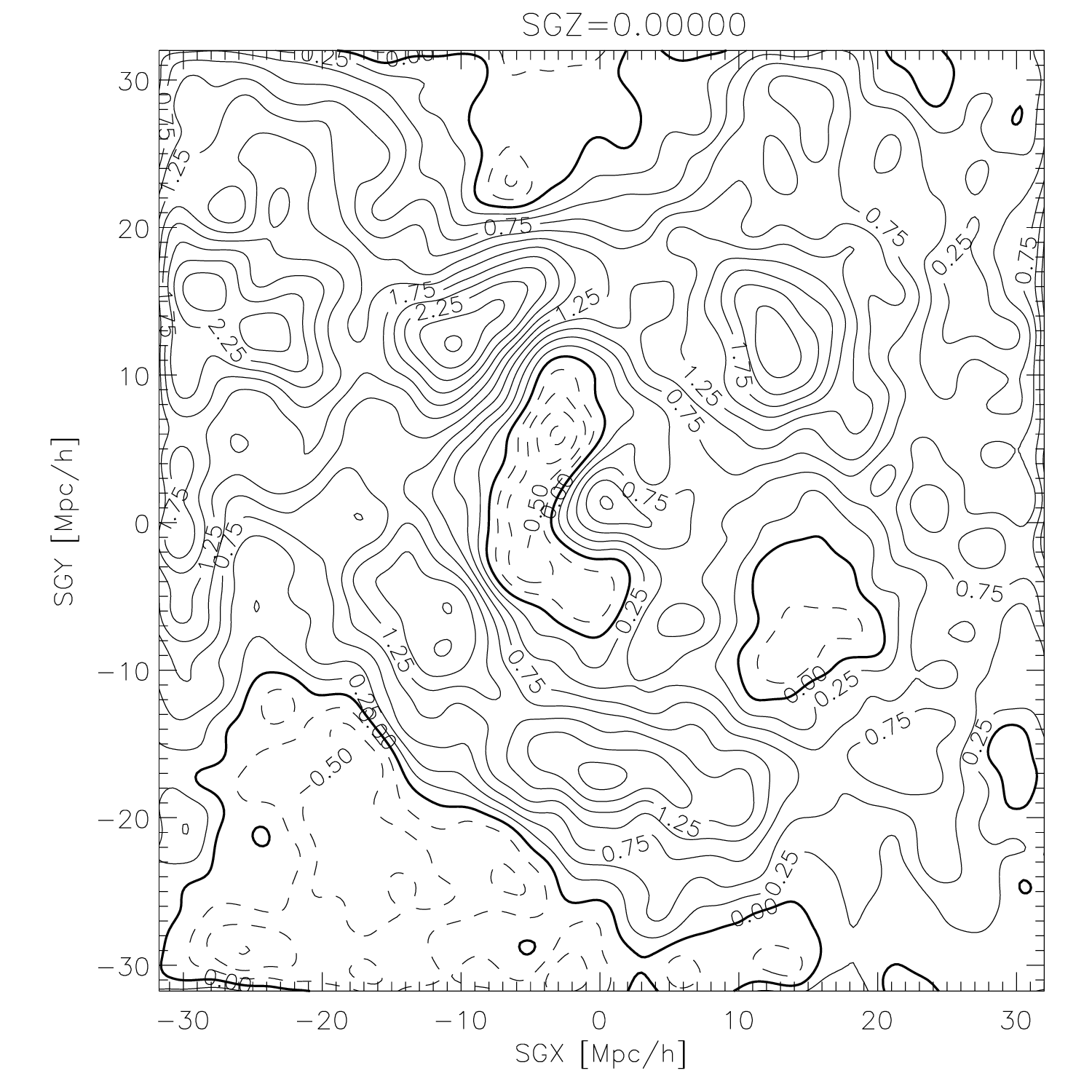}
\includegraphics[width=0.5\textwidth]{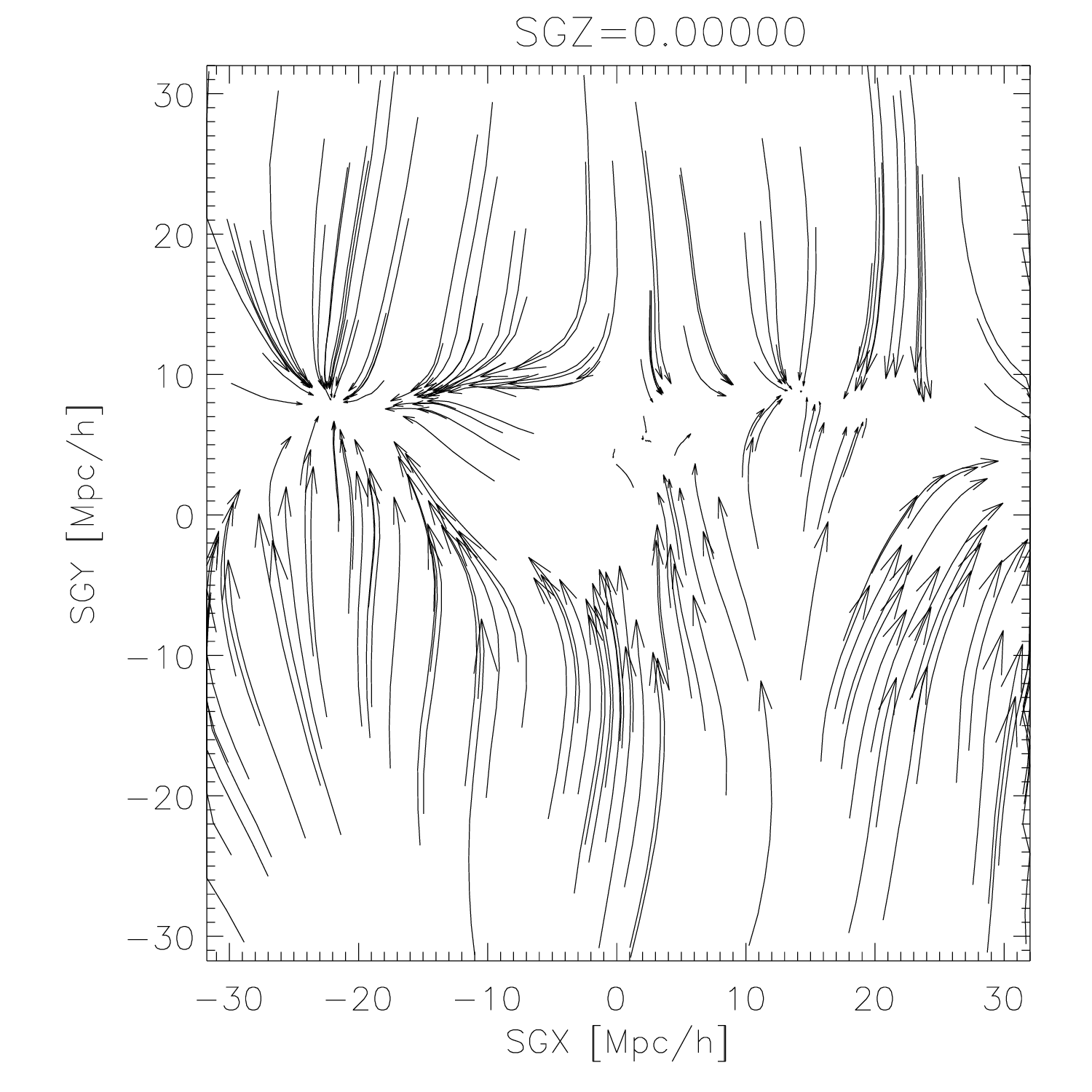}
\caption{The mean linear density and velocity fields. Left: linear density
field smoothed by a Gaussian kernel with $R_g=1\hMpc$, solid, solid
thick and the dashed lines correspond to positive, zero and negative
values. Right: velocity field 
}
\label{fig_h009z:linear}
\end{figure}

Moreover, the ensemble of 200 constrained simulations can be used to
study the statistical properties of the linear constrained
realizations, the non-linear density and velocity fields extracted
from the simulations, and the relation between the constrained initial
conditions and the final simulations. A first step in this direction
is presented here. A sub-ensemble of 60 linear constrained
realizations is studied and the mean and the variance of the linear
density and velocity fields are studied. Fig. \ref{fig_h009z:linear}
presents the mean, taken over the 60 constrained realizations, of the
density (left panel) and the three-dimensional velocity fields. The
presented velocity field corresponds to the divergent component of the
velocity field which is constructed by removing the tidal component
which is produced by the mass distribution outside the computational
box\cite{hoffmanetal:01}.  The cosmography exhibited by the
constrained realizations is dominated by the Local Supercluster,
running at roughly $SGY\approx 15 \hMpc$ across the box, and in
particular by the Virgo cluster at $[SGX, SGY, SGZ] \approx [-8, 15,0]
\hMpc$. Note that linearly recovered Virgo cluster is already shifted
by a few Mpc to the left on the Supergalactic plan. The velocity field
itself converges close to $\approx [-20, 10,0] \hMpc$, and therefore
we expect to see a drift of the simulated Virgo in the negative SGX
direction.

\begin{figure}[t]
\sidecaption[t]
\includegraphics[width=0.5\textwidth]{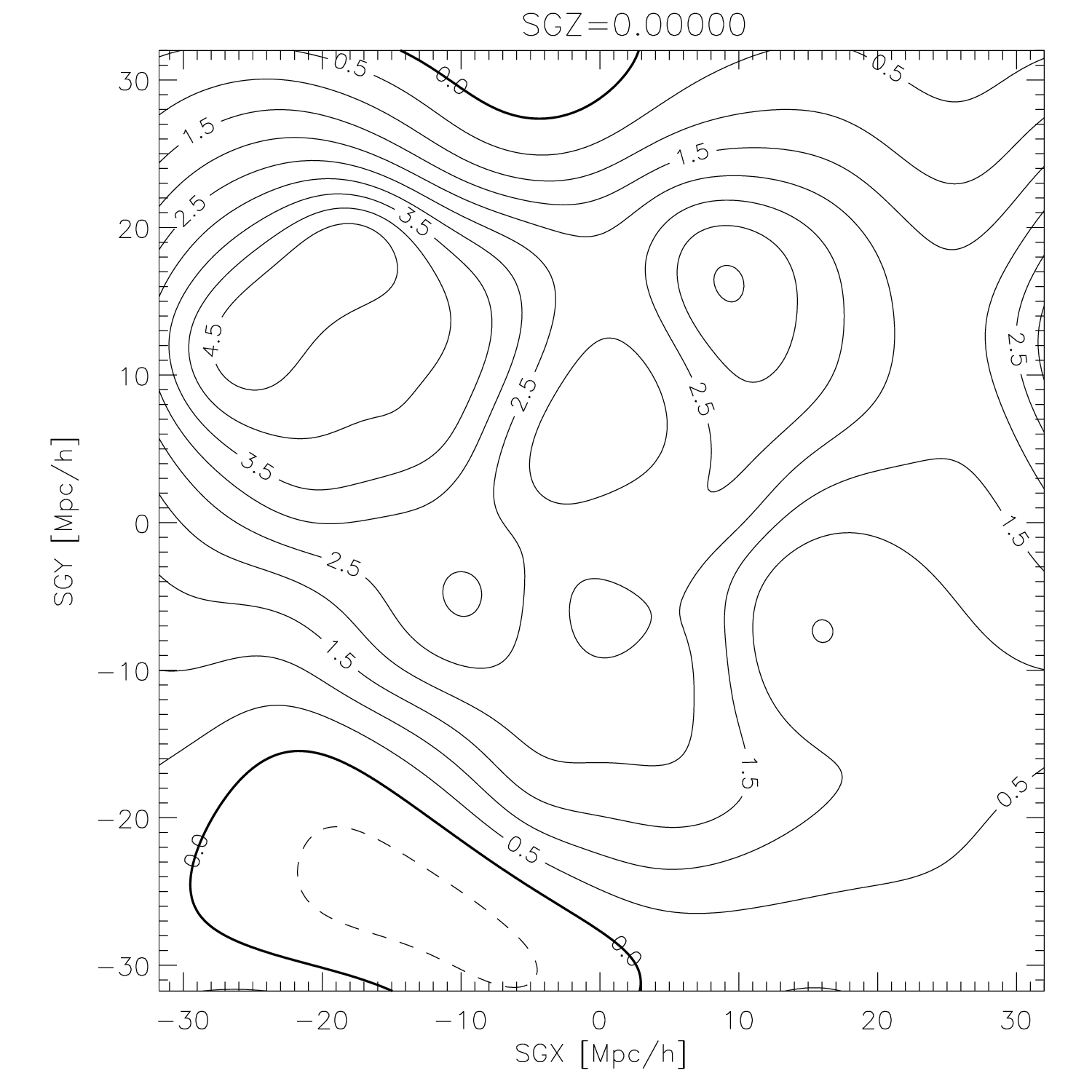}
\includegraphics[width=0.5\textwidth]{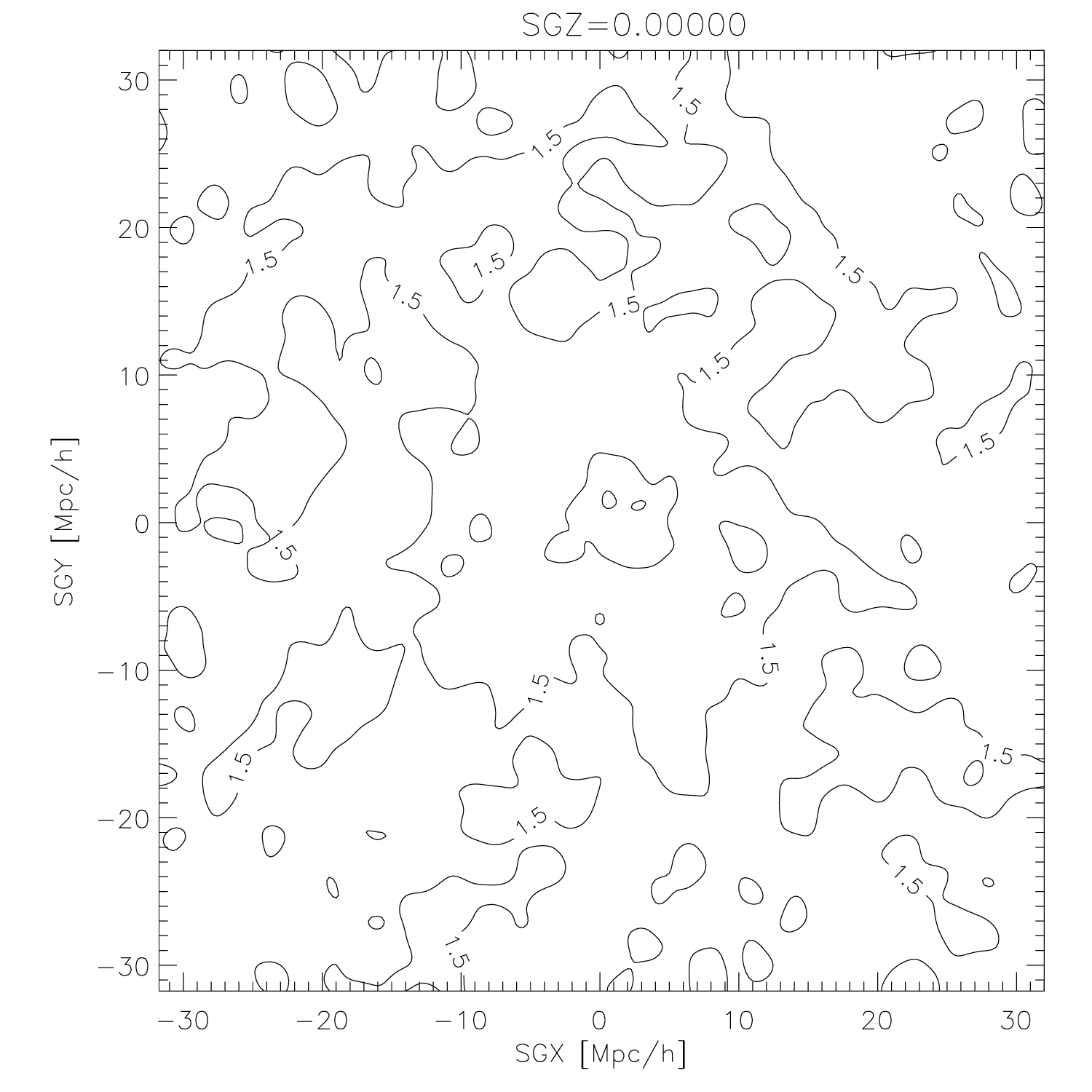}
\caption{Left:The scatter, i.e the standard deviation, around the mean
density field smoothed by a Gaussian kernel with $R_g=5\hMpc$ is
evaluated. The figure shows the mean density field normalised by it
scatter. Right: The same but for $R_g=1\hMpc$.  }
\label{fig_h009z:scatter}
\end{figure}

Fig. \ref{fig_h009z:linear} shows the mean density and velocity
fields, however individual constrained realizations deviate from the
mean. To study the nature of the scatter of the constrained
realizations from the mean field we calculate the scatter, i.e. the
standard deviation, of the density field. It is interesting to see how
robust are the various features uncovered by the constrained
realizations against the scatter exhibited by the constrained
realizations. Fig. \ref{fig_h009z:scatter} presents the mean density
field divided by the local scatter. This normalization of the field by
"sigma" is strongly scale dependent. A Gaussian kernel of
$R_g=5~\hMpc$ is used in the left panel and $R_g=1~\hMpc$ in the right
one.  Normalizing the mean field by the local scatter provides a local
measure of the statistical significance of any feature recovered by
the constrained realizations. The figure shows that indeed the Local
Supercluster is the most robust feature of the constrained
realizations.  Yet, the normalized $R_g=1\hMpc$ map is almost
featureless with typical values of around unity.  It follows that the
$\approx 1\hMpc$ structure is virtually unconstrained by the imposed
data, within the WMAP5 cosmology adopted here, whereas scales larger
than roughly $5\hMpc$ are strongly constrained by the data.

\section{Latest  results from  CLUES simulations}
\label{sec:results_h009z}

Within our project at the Leibniz Rechenzentrum we have performed a
series of constrained simulations with 1 billion particles in boxes of
$160~ \hMpc$ as well as $64~ \hMpc$ side length using both GADGET and
ART. Moreover, we have re-simulated an object which closely resembles
our local group with different resolutions as well as different
physics. In the following two subsections we summarise a few of  the
most interesting  results.

\subsection{Warm Dark Matter in the local Universe}
\label{subsec:WDM_h009z}

We have performed a series of constrained simulations with $1024^3$
dark matter particles assuming the WMAP3 normalization of the power
spectrum\cite{Spergel:07}: $\Omega_m=0.24$, $\Omega_{\Lambda}=0.76$,
$H_0=100h\rm\,km\,s^{-1}Mpc^{-1}$ with $h=0.73$, $n=0.95$ and
$\sigma_8=0.75$.  In order to test the effects of different dark
matter candidates in the structure of the local groups, we also
generated the same initial conditions but assuming that the dark
matter is made of a warm, low mass candidate with a mass
per particle of $m_{\rm WDM}=1\keV$.  The effects of Warm Dark Matter
(WDM) on the structure formation is to remove power from short scales,
due to the large thermal velocities of the particles. For such WDM
particles the free-streaming length is $350\hkpc$ which corresponds to
a filtering mass of $\sim1.1\times10^{10}\hMsun$\cite{Bodeetal01}.
Thus, we just had to change the initial power spectrum, from the
standard CDM, to the WDM which contains a sharp cut-off at the
free-streaming length.  In order to test the effect of the WDM on the
structures formed, we need to have enough mass resolution to resolve
those structures well below the cut-off imposed by the free streaming
of WMD particles.  This translates into a minimum number of particles
in a simulation box such that the Nyquist Frequency imposed by the
discreteness of our sampling of particles, be always smaller than the
frequency at the cut-off in the WDM spectrum.  Using $1024^3$
particles in the simulation box, we ensure that this conditions
fulfils. Thus, we generated the same realization for both CDM and WDM
constrained initial conditions

In a universe filled with Warm Dark Matter the exponential cut-off in
the power spectrum leads naturally to a reduction of small scale
structure. Our choice of $m_{\rm WDM}=1\keV$ is close to the lower
limit for the mass of the WDM particle predicted by observations. We
have chosen this extreme case to study the maximal possible effect of
the Warm Dark Matter on the local structure of the universe as
discussed in more detail in Zavala et al \cite{zavalaetal:09} and
Tikhonov et al \cite{tikhonovetal:09}.  We summarize here the main
results from the analysis of the big simulations with $1024^3$ dark
matter particles assuming the WMAP3 normalization of the power
spectrum\cite{Spergel:07} and two model WDM and CDM that were
described in the previous sections.

In the $\Lambda$CDM case the mass function of halos in the whole
simulated box follows closely the estimates from the Sheth \& Tormen
formalism, except in the highest mass end due to the influence of the
Local Supercluster, a massive structure in this small box which comes
in due to the observational constraints. In the Warm Dark Matter
universe the mass function lies close to the CDM one for halos with
masses higher than the filtering mass, however for lower masses it
flattens and then rise due to spurious numerical fragmentation for
masses ${\lesssim} 3\times10^9\hMsun$ \cite{wangwhite:07}.  The mass
resolution of $m_{\rm DM}=1.63\times10^7\hMsun$, allows us to derive
robust results for halos with masses larger than this limiting mass,
corresponding to maximum rotation velocities of 24$\kms$.

\begin{figure}[t]
\sidecaption[t]
\includegraphics[width=.8\textwidth]{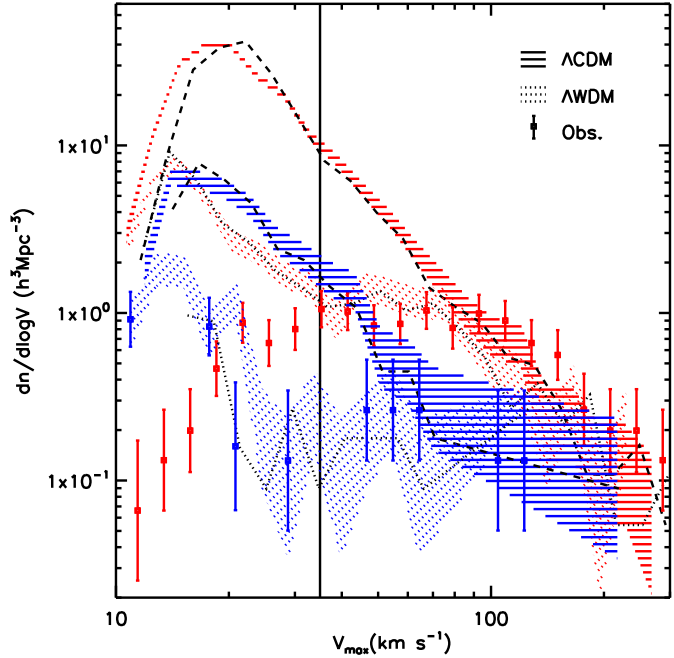}
\caption{Velocity function for the sample of galaxies in the
Virgo-direction region taken taken from the ALFALFA catalogs (red
square symbols with error bars).  Predictions from our constrained
simulations for the observed field of view appear as the dashed
($\Lambda$CDM) and dotted ($\Lambda$WDM) red areas, delimited by
Poisson error bars.  In blue the same is shown for the
anti-Virgo-direction.  The vertical solid line marks the value of
$V_{max}$ down to which the simulations and observations are both
complete.  
}  
\label{vel_func} 
\end{figure}

From the Warm and Cold Dark Matter simulations we have calculated the
velocity functions of haloes which has the advantage over the mass
function that it can be compared more directly with observational
data. From our simulations we have obtained predictions for the
velocity function of halos in the field of view, within 20$\hMpc$,
that is being surveyed by the ongoing Arecibo Legacy Fast ALFA
(ALFALFA) HI blind survey \cite{giovanelli:05,giovanelli:07}.  This
survey is divided into two regions on the sky, one includes the Virgo
cluster, the other points into the opposite direction.  Additionally
constrained to distances less than $20\hMpc$ we call these here the
``Virgo-direction region'' and ``anti-Virgo-direction region'',
respectively. Zavala et al. \cite{zavalaetal:09} compared the
constrained simulations with the ALFALFA observational data. The
results are summarised in Fig.\ref{vel_func}.

For velocities in the range between $80\kms$ and $300\kms$, the
velocity functions predicted for the $\Lambda$CDM and $\Lambda$WDM
simulations agree quite well with the observed velocity functions.
This result is encouraging because it confirms that our constrained
simulations properly reflect our local environment. In fact, the
simulations are able to predict the shape and normalization of the
velocity function in the high velocity regime, in particular in the
Virgo direction, where the normalization is an order of magnitude
larger than for the universal velocity function.

The minimum mass for which we can trust the simulations and
observations corresponds to a maximum circular velocity of $\sim35\kms$
(solid line in Fig. \ref{vel_func}). Between this velocity
and about $V_{max}\sim80\kms$ the predictions agree well for the
$\Lambda$WDM cosmogony. On the contrary, the $\Lambda$CDM model
predicts a steep rise in the velocity function towards low
velocities. Thus, ar our limiting circular velocity,
$V_{max}\sim35\kms$, it forecasts $\sim10$ times more sources both in
Virgo-direction (red) as well as in anti-Virgo-direction (blue) than
the ones observed by the ALFALFA survey. These results indicate a
potential problem for the cold dark matter paradigm
\cite{zavalaetal:09}. The spectrum of mini-voids also points to a
potential problem of the $\Lambda$CDM
model\cite{klypintikhonov:09}. The $\Lambda$WDM model provides a
natural solution to this problem, however, the late formation of halos
in the $\Lambda$WDM model might be a problem for galaxy
formation\cite{tikhonovetal:09}.

\subsection{Satellites in the Local Group}
\label{subsec:sat_h009z}

\begin{figure}[t]
\sidecaption[t]
\includegraphics[width=\textwidth]{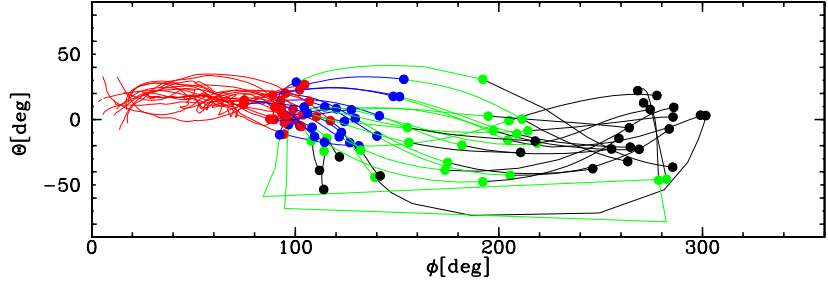}
\caption{Orbits of the large in-falling group of sub-haloes as seen by
the observer placed at the centre of M31. The spherical coordinate
system was aligned to match the angular momentum vector of the host
which is pointing along the $z$ axis. Different colours correspond to
different times. Red circles: 3 Gyrs ago, blue: 2 Gyrs ago, green: 1
Gyr ago, black: present time.
} 
\label{fig_h009z:LIG}
\end{figure}

Klimentowksi et al.  used the dark matter only zoomed simulation of the Local
Group with an effective  resolution of  $4096^3$ particles as a numerical
laboratory for studying the evolution of the population of its
sub-haloes\cite{klimentowskietal:09}. In M31 they have detected a large
group of in-falling satellites which consists of 30
haloes. Fig.~\ref{fig_h009z:LIG} shows the orbits of the satellites
projected on to the sky. The observer was placed in the centre of M31
and the coordinate system was chosen so that the $z$-axis is aligned
with the angular momentum of M31. About 4 Gyrs ago the satellites fell
into M31 from one well-defined region in the sky but the spread
increases with time.  This example suggests that the alignment of
angular moments of in-falling satellites is not well conserved, even
though the group has not yet decayed. One should thus be very careful
when trying to reproduce the histories of dwarf galaxies using their
present proper motions.

\begin{figure}[t]
\sidecaption[t]
\includegraphics[width=0.5\textwidth]{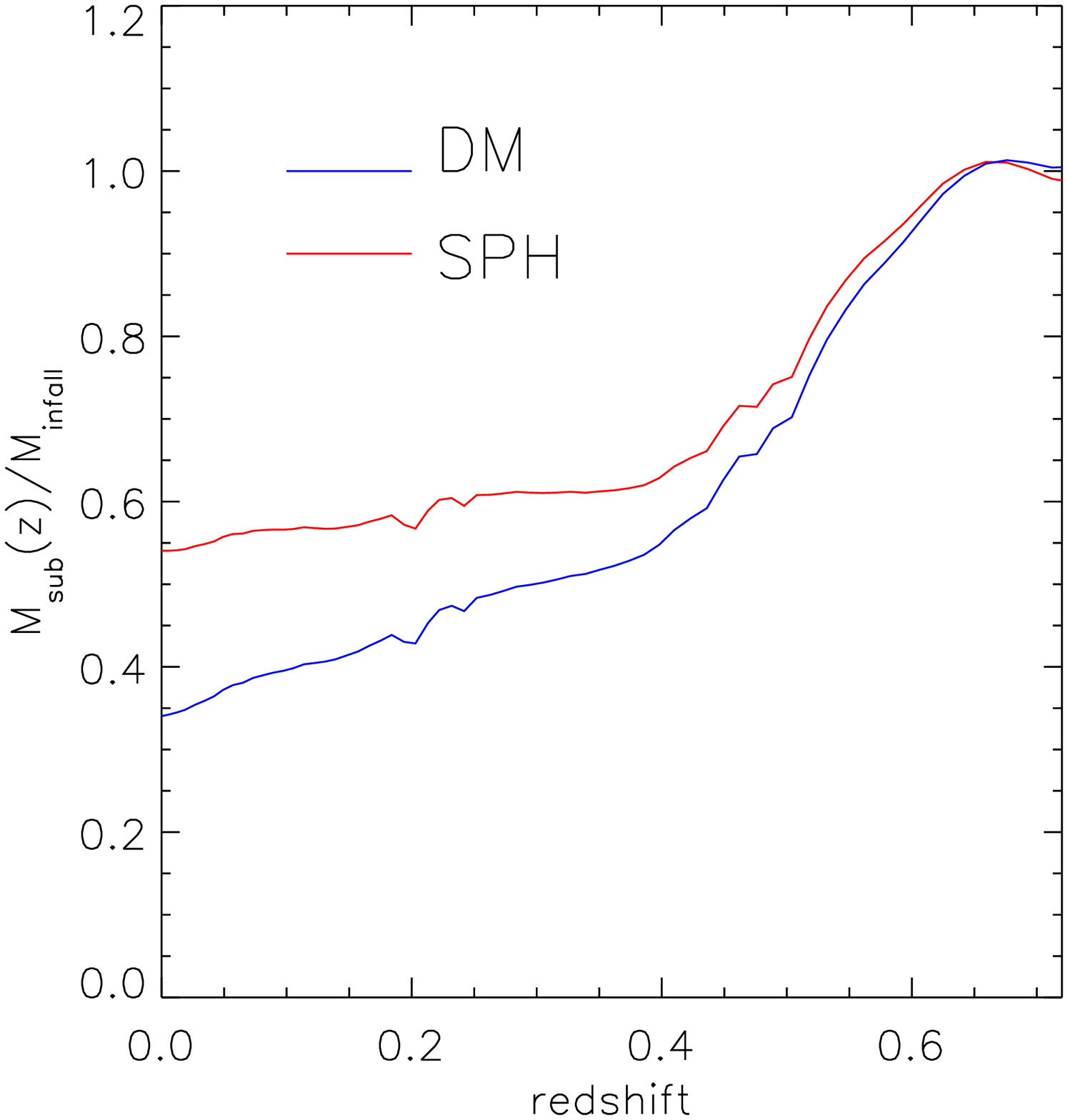}
\includegraphics[width=0.5\textwidth]{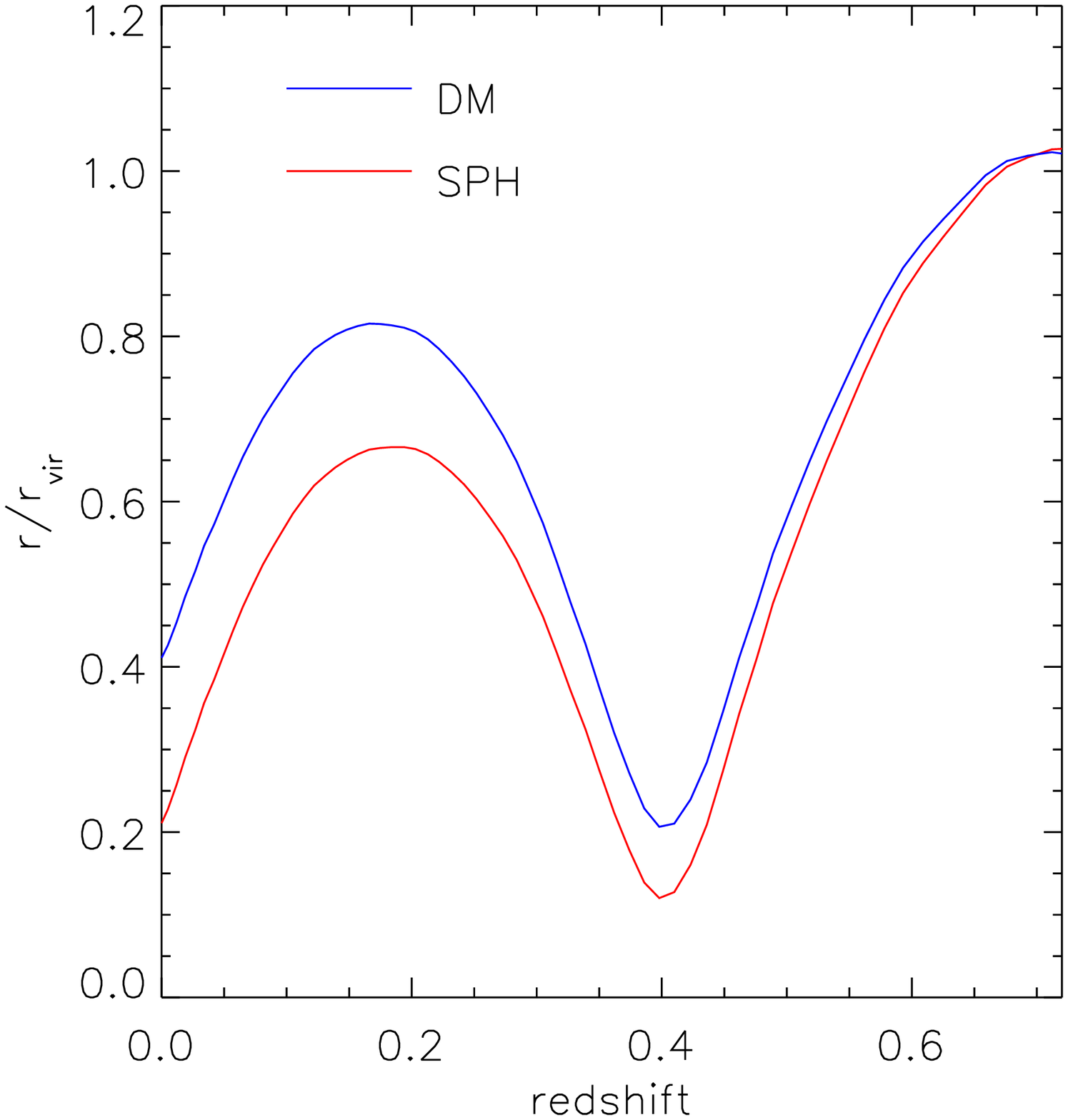}
\caption{Left: The mass of an SPH sub-halo (red line) and its DM sister
(blue line) as a function of redshift, normalised to the mass each
sub-halo had at infall, around $z=0.72$. Right: The distance from the
main halo of an SPH sub-halo and its DM sister as a function of
redshift.
} 
\label{fig_h009z:SPH_DM}
\end{figure}

Since we were running simulations of the LG  with and without
baryons modelled hydrodynamically we can quantify the effect of gas
physics on the $z=0$ population of sub-haloes. We found
\cite{libeskindetal:09} that above a certain mass cut, $M_{\rm sub} >
2\times10^{8}h^{-1} M_{\odot}$ sub-haloes in gas-dynamic simulations are
more radially concentrated than those in in dark matter only
simulations.  The increased central density of such sub-halos results
in less mass loss due to tidal stripping than the same sub-halo
simulated with only dark matter. The increased mass in hydrodynamic
sub-haloes with respect to dark matter ones causes dynamical friction
to be more effective, dragging the sub-halo towards the centre of the
host. In Fig. \ref{fig_h009z:SPH_DM} we show as an example the mass
and radial position of an in-falling satellite as a function of
time. The same satellite has been identified in the dark matter only
simulation (blue) and the gas-dynamical simulation (red).  In the left
panel one can see that the sub-halo's mass in the dark matter only
simulation closely follows the SPH sub-halo's mass for a short period
directly after accretion. However, by $z=0.4$ the DM halo has already
lost a good $\sim10\%$ more then its SPH counterpart. By $z=0$ the DM
sister has lost $\sim65\%$ of its infall mass, while the SPH sub-halo
has only lost $\sim 45\%$ of its infall mass. In the right panel we
show the orbit of this sub-halo as a function of redshift which diverge
shortly after infall. By the first pericentric passage, the SPH
sub-halo penetrates further in by about a factor of 2 than the DM
sister. The sub-halo from the dark matter only simulation is
consistently further behind the SPH sub-halo at all stages of the
orbit, resulting in the SPH sub-halo being nearly a factor of 2 closer
to the host halo's centre then the DM sub-halo at $z=0$.

\section{Summary}
\label{sec:summary_h009z}

Within the CLUES project ({\tt http://clues-project.org}) we have
performed a series of constrained simulations of the local
universe. These simulations reproduce the observed large scale
structures around the Local Group whereas small scale structures below
scales of $\sim 1 \hMpc$ are essentially random. In four simulations
we have identified Local Groups situated at the right position but
having different merging histories. Constrained simulations are a
useful tool to study the formation and evolution of our Local Group in
the right cosmological environment and the best possible way to make a
direct comparison between numerical results and observations,
minimizing the effects of the cosmic variance.

\begin{acknowledgement} The computer simulations described here have
been performed at LRZ Munich and BSC Barcelona.  We acknowledge
support by ASTROSIM, DFG, DEISA and MICINN (Spain) for our
collaboration.  We thank Jesus Zavala (Munich), Noam Libeskind
(Potsdam), Jarek Klimentowski (Warsaw) and Kristin Riebe (Potsdam) for
providing plots from our common research projects.

\end{acknowledgement}

\end{document}